\documentclass[floatfix,twocolumn,aps]{revtex4}
\usepackage{graphicx,float}

\begin{document}

\title{The Network Topology of the Interbank Market 
\footnote{
The views expressed in this paper are 
stritcly the views of the authors and 
do in no way commit the OeNB. \\
{\bf Correspondence to:}
Stefan Thurner;  Complex Systems Research Group, 
HNO, AKH-Wien, University of Vienna; 
W\"ahringer G\"urtel 18-20; A-1090 Vienna, Austria; 
Tel.: ++43 1 40400 2099;  Fax: ++43 1 40400 3332; 
e-mail: thurner@univie.ac.at
}
}

\author{
Michael Boss$^{1}$, 
Helmut Elsinger$^{2}$, 
Martin Summer$^{1}$, and 
Stefan Thurner$^{3}$ \\
\vspace{0.4cm}
 $^{1}${\it Oesterreichische Nationalbank,  
  Otto-Wagner-Platz 3, A-1011 Wien; Austria}  \\
 $^{2}${\it Department of Finance, Universit\"at Wien, 
  Br\"{u}nner Strasse 71, A-1210 Wien; Austria} \\  
 $^{3}${\it Complex Systems Research Group, HNO, 
  Universit\"at Wien, W\"ahringer G\"urtel 18-20, A-1090; Austria} \\
}
\begin{abstract}
We provide an empirical analysis of the network structure 
of the Austrian interbank market based on 
a unique data set of the Oesterreichische Nationalbank (OeNB). 
We show that the contract size 
distribution follows a power law over more than 3 decades.  
By using a  novel ''dissimilarity'' measure 
we find that the interbank network shows a community structure 
that exactly mirrors the regional and sectoral organization of the 
actual Austrian banking system. 
The degree distribution of the 
interbank network shows two different power law exponents which 
are one-to-one related to two sub-network structures, differing in 
the degree of hierarchical organization. 
The banking network moreover shares typical structural features  
known in numerous complex real world networks:
a low clustering coefficient  
and a relatively short average shortest path length.
These empirical findings are in marked contrast to 
interbank networks that have been analyzed in 
the theoretical economic and econo-physics literature.  
\end{abstract}
 
\maketitle

Over the past years the physics community has largely contributed 
to the analysis and to a functional understanding of the structure of 
complex real world networks.  
A key insight of this research has been the discovery of surprising 
structural similarities in very different networks, such as 
Internet, the World Wide Web, 
collaboration networks, biological networks, 
communication networks, electronic circuits and power grids, 
industrial networks, nets of software components and
energy landscapes. See \cite{dormen03} for an overview. 
Remarkably, most of these real world networks 
show  degree distributions which   
follow a power law, feature a certain pattern of cliquishness, 
quantified by the  {\it clustering coefficient} of the network, 
and exhibit the so 
called {\it small world phenomenon}, meaning that  
the average shortest path between any two vertices 
(''degrees of separation'') in the network can be surprisingly small 
\cite{watts}. 
Maybe one of the most important contributions to recent network 
theory is an interpretation of these network parameters with respect 
to stability, robustness, and efficiency of an underlying system, 
e.g. \cite{barabasi00}. 

From this perspective  financial networks seem a natural 
candidate to study. Indeed in the economic literature
financial crises that have hit countries all around the globe
have led to a boom of papers on banking-crises, 
financial risk-analysis and to  numerous policy 
initiatives to improve financial stability. 
One of the major concerns in these debates is the danger of so 
called {\it systemic risk}: the large scale breakdown of financial 
intermediation due to domino effects of insolvency \cite{hardeb00, sum03}.
The network of mutual credit relations between 
financial institutions is supposed to play a key role
in the risk for contagious defaults. 
Some authors in the theoretical economic literature
on contagion \cite{algal00,freparroc00,thurner03} suggest
network topologies that might be interesting to look at.
In \cite{algal00}  it is suggested to study a complete
graph of mutual liabilities. The properties of a banking system with
this structure is then  compared to properties of 
systems with non complete networks.
In \cite{freparroc00} a circular graph is contrasted with a 
complete graph.  In \cite{thurner03} a 
much richer set of different network structures is studied.
Yet, surprisingly little is known 
about the {\it actual} empirical network topology of mutual
credit relations between financial institutions. To our  best 
knowledge the network topology 
of interbank markets has so far not been studied empirically. 


The interbank network is characterized by the 
liability (or exposure)  matrix $L$. The entries 
$L_{ij}$ are the liabilities bank $i$ 
has towards bank $j$.  We use the convention to write liabilities
in the rows of $L$. If the matrix is read column-wise 
(transposed matrix $L^T$) we see the claims or interbank 
assets,  banks hold with each other. 
Note, that $L$ is a square matrix but not necessarily symmetric. 
The diagonal of $L$ is zero, i.e. no bank self-interaction exists.
In the following we are looking for the bilateral 
liability  matrix $L$ of all (about $N=900$) Austrian banks, 
the Central Bank (OeNB) and an aggregated foreign banking 
sector. Our data consists of 10 $L$ matrices each representing 
liabilities for quarterly single month periods between 
$2000$ and $2003$.   
To obtain  the Austrian interbank network from Central Bank   
data we draw upon two major sources: 
we exploit structural features of the Austrian bank 
balance sheet data base (MAUS) and the 
major loan register (GKE) in combination with an estimation technique. 

The Austrian  banking system has a sectoral organization 
due to historic reasons. Banks belong to one of seven sectors: 
savings banks (S),
Raiffeisen (agricultural) banks (R), 
Volksbanken (VB), 
joint stock banks (JS), 
state mortgage banks (SM), 
housing construction savings and loan associations (HCL), and 
special purpose banks (SP). 
Banks have to break down their balance sheet
reports on claims and liabilities with other banks according 
to the different banking sectors, Central Bank and 
foreign banks. This practice of reporting on balance interbank
positions breaks the liability matrix $L$ down to blocks of sub-matrices for
the individual sectors. 
The savings banks and the Volksbanken sector are organized 
in a two tier structure with a sectoral head institution. 
The Raiffeisen sector is organized by a three tier structure,
with a head institution for every federal state of Austria.
The federal state head institutions have a central institution,
Raiffeisenzentralbank (RZB) which is at the top of the Raiffeisen
structure. Banks with a head institution have to disclose their
positions with the head institution, which gives additional 
information on $L$. Since many banks in the system hold interbank 
liabilities only with their head institutions,  one 
can pin down many entries in the $L$ matrix exactly. 
This information is combined in a next step with the data 
from the major loans register of OeNB. This 
register contains all interbank loans above a threshold of 360 000 Euro. 
This information provides us with a set of constraints 
(inequalities) and zero restrictions for individual entries $L_{ij}$.
Up to this point one can obtain about 
$90\%$ of the $L$-matrix entries exactly. 

\begin{figure}[htb]
\begin{center}
\begin{tabular}{c} 
\includegraphics[height=60mm]{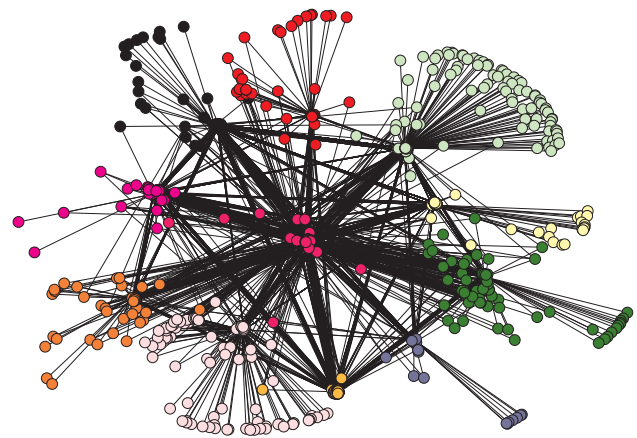} \\
{\large (a)} \\
\includegraphics[height=65mm]{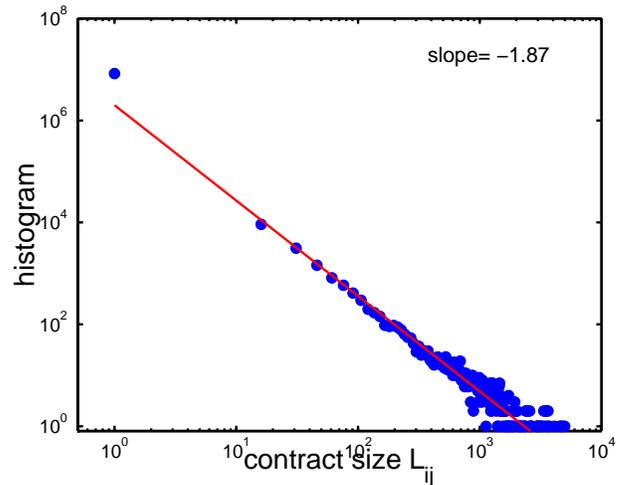} \\
{\large (b)} \\
\end{tabular}
\end{center}
\caption{
The banking network of Austria (a). Clusters are grouped (colored) according 
to regional and sectorial organization: 
R-sector with its federal state sub-structure: 
RB yellow,  RSt orange, light orange RK, 
gray RV, dark green RT, black RN, light green RO,
light yellow RS.
VB-sector: dark gray, S-sector: orange-brown, other: pink.
Data is from the September 2002 $L$ matrix, which is representative for 
all the other matrices. 
In (b) we show the contract size distribution within this network 
(histogram of all entries in $L$) which follows a power 
law with exponent $-1.87$. 
Data is aggregated from all 
10 matrices. 
}
\label{pajek} 
\end{figure}
For the rest we employ an estimation routine  
based on local entropy maximization,  
which has already been used to  
reconstruct unknown bilateral interbank exposures 
from aggregate information \cite{uppwor02,shemau98}. 
The procedure finds a matrix that fulfills all the known constraints  
and treats all other parts (unknown entries in $L$) 
as contributing equally to the known row and column sums. 
These sums are known since the total claims to other banks 
have to reported to the Central Bank.
The estimation problem can be set up as follows:
Assume we have a total of $K$ constraints. 
The column and row constraints take the form 
\begin{equation}
\sum_{j=1}^{N}L_{ij}=b_i^r \quad \forall \quad i \quad \text{and} \quad 
\sum_{i=1}^{N}L_{ij}=b_j^c \quad \forall \quad j
\label{equcon}
\end{equation}
with $r$ denoting {\it row} and $c$ denoting {\it column}. 
Constraints imposed by the knowledge about particular entries in $L_{ij}$ 
are given by
\begin{equation}
b^l \leq L_{ij} \leq b^u \quad \text{for some $i,j$.}
\label{inequcon}
\end{equation}
The aim is to find the matrix $L$ 
(among all the matrices fulfilling the constraints) 
that has the least discrepancy to some a priori matrix $U$ with respect 
to the (generalized) cross entropy measure 
\begin{equation}
{\cal C}(L,U)=\sum_{i=1}^{N}\sum_{j=1}^{N}L_{ij}\ln 
\left(\frac{L_{ij}}{U_{ij}}\right)  \quad . 
\label{zielf}
\end{equation}
$U$ is the matrix which contains all known exact liability 
entries. For  those entries (bank pairs) ${ij}$ where we have 
no knowledge from Central Bank data, we set $U_{ij}=1$. 
We use the convention that $L_{ij}=0$ whenever $U_{ij}=0$ and 
define $0\ln (\frac{0}{0})$ to be $0$. 
This is a standard convex optimization problem, the 
necessary optimality conditions can be solved efficiently 
by an algorithm described in \cite{fanrajtsa97,bligra97}. 
As a result we obtain a rather precise (see below) 
picture of the interbank relations at a particular point in time. 
Given $L$ we plot the distribution (pdf) of its entries in 
Fig. \ref{pajek}(b).  The distribution of liabilities 
follows a power law for more than three decades with an 
exponent of $-1.87$, which is within a range which is 
well known from wealth- or firm size distributions 
\cite{solomon,axtell}.

To extract the network topology from these data, there are three 
possible approaches to describe the structure as a graph. 
The first approach is to look at the liability matrix as a 
{\it directed graph}.
The vertices are all Austrian banks. The Central Bank OeNB, and the 
aggregate foreign banking sector are represented by a single vertex each. 
The set of all initial (starting) vertices is the set of banks 
with liabilities in the interbank market;  
the set of end vertices is the set of all banks that are 
claimants in the interbank market. 
Therefore, each bank that has liabilities with some other 
bank in the network is considered an initial vertex in the 
directed liability graph. Each bank for which this liability 
constitutes a claim, i.e. the bank acting as a counterparty,
is considered an end vertex in the directed liability graph.
We call this representation the {\it  liability adjacency 
matrix} and denote it by $A^l$ ($^l$ indicating liability).  
$A^l_{ij}=1$ whenever a connection starts 
from row node $i$ and leads to column node $j$, and $A^l_{ij}=0$ 
otherwise. 
If we take the transpose of $A^{l}$ we get  
the interbank asset matrix $A^{a}=(A^{l})^T$.
A second way to look at the graph is to ignore directions and 
regard any two banks as connected if they
have either a liability or a claim against each other. 
This representation results 
in an undirected graph whose corresponding adjacency 
matrix $A_{ij}=1$ whenever we observe an
interbank liability or claim. 
Our third graph representation is to define an undirected but weighted 
adjacency matrix $A^w_{ij}=L_{ij}+L_{ji}$, which measures the gross 
interbank interaction, i.e. the total volume of 
liabilities and assets for each node. 
Which representation to use depends on the questions addressed  
to the network. For statistical descriptions of the network structure 
matrices $A$, $A^a$, and $A^l$ will be sufficient, to reconstruct the 
community structure from a graph, the weighted 
adjacency matrix $A^w$ will be the more useful choice. 


There exist various ways to find functional clusters 
within a given network. Many algorithms take 
into account local information around a given vertex, 
such as the number of nearest neighbors shared with other 
vertices, number of paths to other vertices, see e.g.
\cite{wasserman94, ravasz02}. 
Recently a global algorithm was suggested which 
extends the concept of vertex betweenness \cite{freeman77}
to links \cite{girvan02}. This elegant 
algorithm outperforms most traditional approaches 
in terms of mis-specifications of vertices to clusters, 
however it does not provide 
a measure for the differences of  clusters.  
In \cite{zhou03_b} an algorithm was introduced 
which - while having at least the same performance 
rates as \cite{girvan02} - provides such a measure, 
the so-called dissimilarity index. The algorithm is based on a 
distance definition presented in \cite{zhou03_a}. 

\begin{figure}[htb]
\begin{center}
\includegraphics[height=60mm]{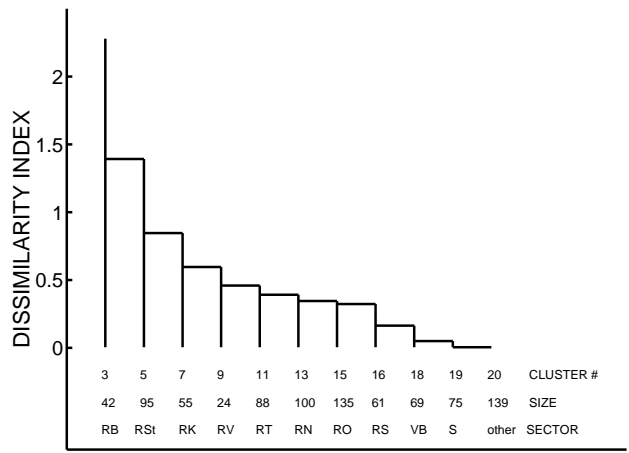}
\end{center}
\caption{
Community structure of the Austrian interbank market 
network from the September 2002 data. The dissimilarity 
index is a measure of the ''differentness'' of the clusters.}
\label{community} 
\end{figure}

For analyzing our interbank network we apply this latter 
algorithm to the weighted adjacency matrix $A^w$. 
As the only preprocessing step we clip all entries in 
$A^w$ above a level of 300 m Euro for numerical reasons, i.e. 
$A^w_{\rm clip} = \min(A^w,300 {\rm m})$. 
The  community structure obtained in this way (Fig. \ref{pajek}a) 
can be compared to the actual community structure in the real 
world.  
The result for the community structure obtained from 
one representative data set is shown in 
Fig. \ref{community}. Results from other datasets are practically 
identical. 
The algorithm identifies communities of banks which are 
coupled by a two or three tier structure, i.e. the 
R, VB, and S sectors. 
For banks which in reality are not structured in a hierarchical 
way, such as  banks in the SP, JS, SM, HCL sectors, no strong community 
structure is expected. By the algorithm these banks are 
grouped together in a cluster called 'other'. 
The Raiffeisen sector, with its substructure in federal 
states, is further grouped into clusters which are clearly identified 
as R banks within one of the eight federal states (B,St,K,V,T,N,O,S). 
In Fig. \ref{community} these clusters are marked as e.g. 'RS', 
'R' indicating the Raiffeisen sector, and 'S' the state of 
Salzburg. 
Overall, there were 31 mis-specifications into wrong clusters within
the total $N=883$ banks, which is a mis-specification rate of 3.5 \%, 
demonstrating the quality of the dissimilarity algorithm and - more 
importantly - the quality of the entropy approach to reconstruct 
matrix $L$. 


\begin{figure}[htb]
\begin{center}
\includegraphics[height=60mm]{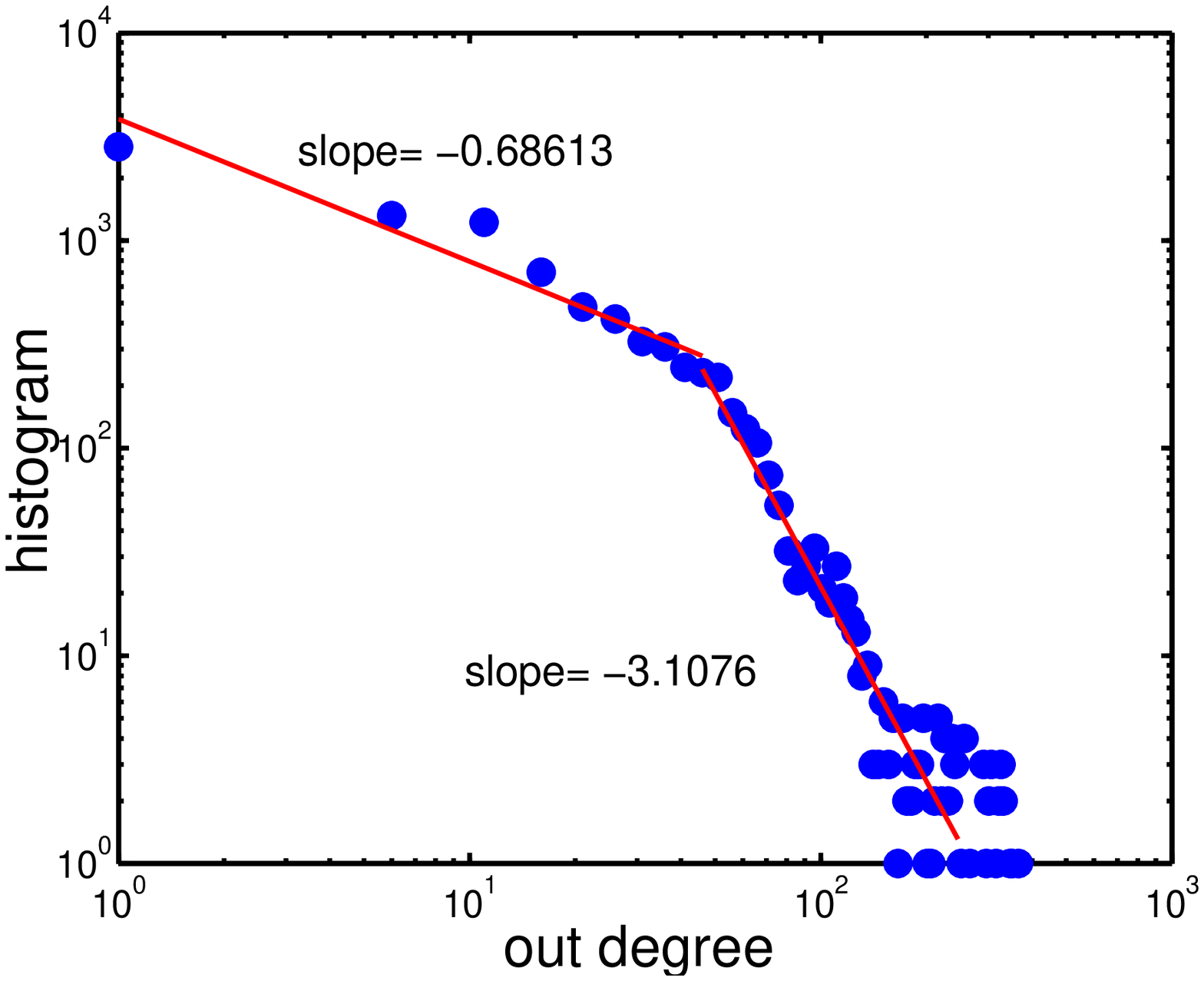}\\
\vspace{-2cm}
\hspace{-5cm} {\large (a)} \\
\vspace{1.5cm}
\includegraphics[height=60mm]{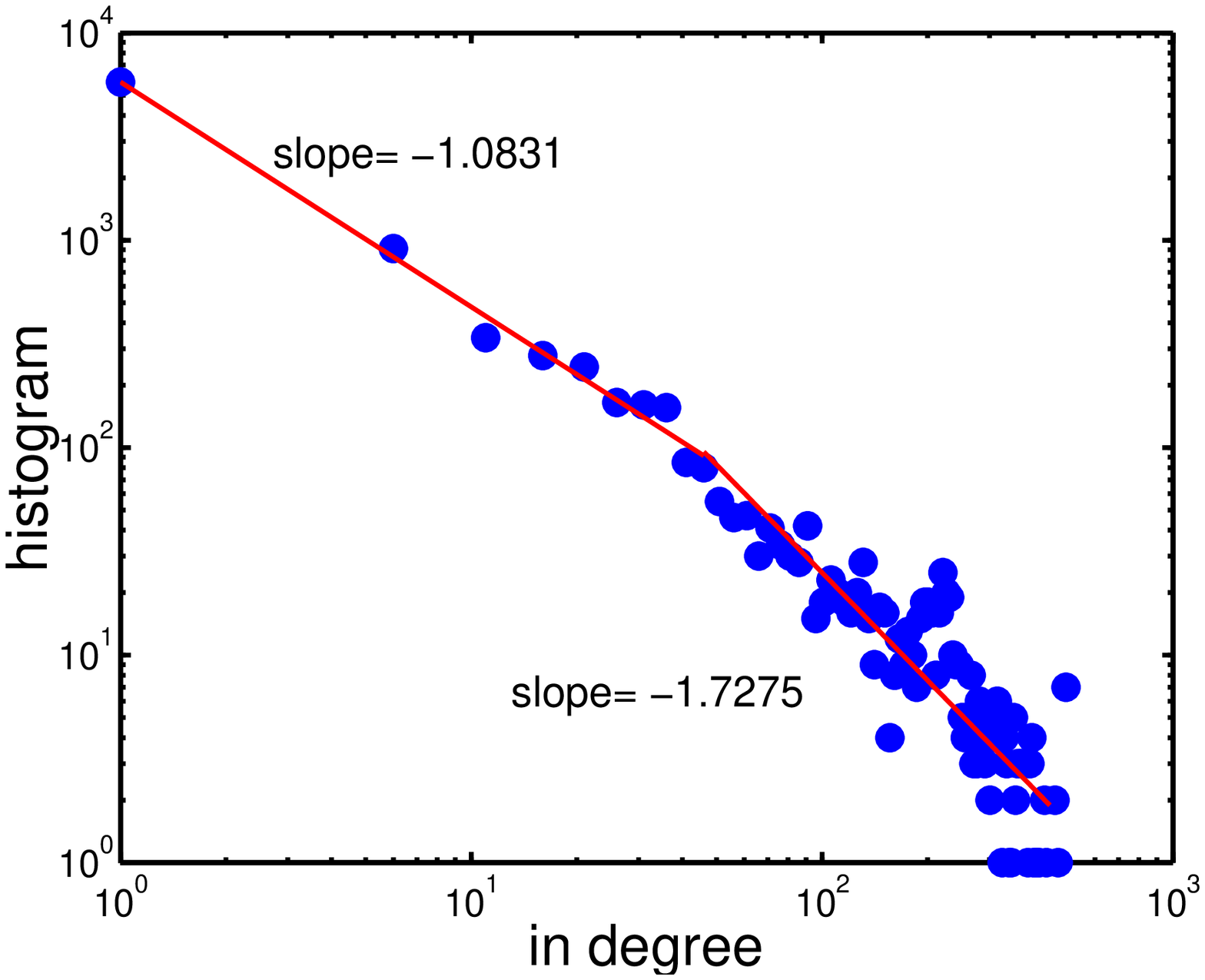}\\
\vspace{-2cm}
\hspace{-5cm} {\large (b)} \\
\vspace{1.5cm}
\includegraphics[height=60mm]{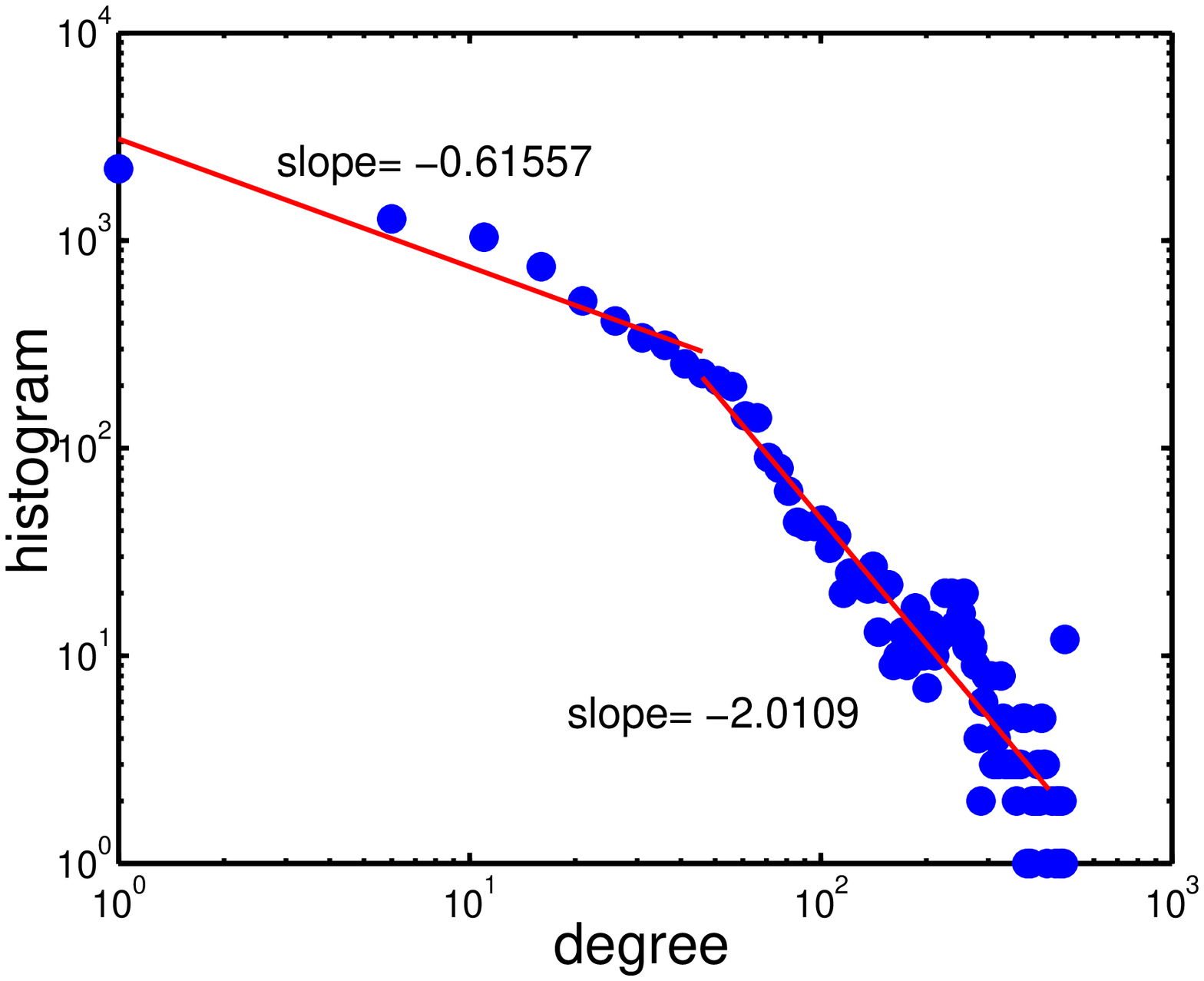}\\
\vspace{-2cm}
\hspace{-5cm} {\large (c)} \\
\vspace{1.5cm}
\end{center}
\caption{
Empirical out-degree (a) in-degree and (b) distribution of the interbank 
liability network. In (c) the degree distribution of the interbank 
connection network is shown. All the plots are histograms 
of aggregated data from all the 10 datasets.}
\label{graph} 
\end{figure}

\noindent
{\bf Degree Distributions:}
Like many real world networks, the degree distribution 
of the interbank market follow power laws for 
all three graphs $A^l$, $A^a$, and $A$. 
Figure \ref{graph} (a) and  (b) show the out-degree (liabilities) 
and in-degree (assets) 
distribution of the vertices in the interbank liability network. 
Figure \ref{graph} (c) shows degree 
distribution of the interbank connection graph $A$.  
In all three cases we find two regions which can be fitted 
by a power. Accordingly, we 
fit one regression line to the small degree distribution 
and one to the obvious power tails of 
the data using an iteratively re-weighted least square algorithm. 
The power decay exponents $\gamma_{tail}$ to the tails of the 
degree distributions 
are $\gamma_{tail}(A^l)=3.11$, $\gamma_{tail}(A^a)=1.73$, 
and $\gamma_{tail}(A)=2.01$.
The size of the out degree exponent is within the range of 
several other complex networks, like e.g. 
collaboration networks of actors (3.1) \cite{albert2000}, 
sexual contacts (3.4) \cite{liljeros2001}.
Exponents in the range of 2 are  for example the 
Web (2.1) \cite{albert99} or mathematicians' collaboration
networks (2.1) \cite{barabasi2002}, and examples for exponents 
of about 1.5 are email networks (1.5) \cite{ebel2002}
and co-autorships (1.2) \cite{newman01}.
For the left part of the distribution (small degrees) we
find  $\gamma_{small}(A^l)=0.69$, $\gamma_{small}(A^a)=1.01$, 
and $\gamma_{small}(A)=0.62$. 
These exponents are small compared to 
other real world networks. Compare e.g. 
with foodwebs (1) \cite{montoya2001}. 
We have checked that the distribution for the low degrees 
is almost entirely dominated by banks of the R sector. 
Typically in the R sector most small agricultural banks have links to their
federal state head institution and very few contacts with other banks, 
leading to a strong hierarchical structure, which is also visible 
by plain eye in Fig. \ref{pajek}a. This hierarchical 
structure is perfectly reflected by the 
small scaling exponents \cite{trusina2003}.  

\noindent
{\bf Clustering Coefficient:}
To quantify clustering phenomena within the banking  network 
we use the so-called clustering coefficient $C$, defined by
\begin{equation}
C=\frac{3\times \text{(number of triangles on the graph)}}
{\text{number of connected triples of vertices}} 
\quad . 
\end{equation}
It provides the probability that two vertices that are connected to  any 
given vertex are also connected with one another. 
A high clustering coefficient means that two banks that have interbank 
connections with a third bank, have a greater probability to have interbank 
connections with one another, than will any two banks 
randomly chosen on the network. The clustering coefficient
is only well defined in undirected graphs. 
We find the clustering coefficient of the  
connection network ($A$) to be  
$C=0.12\pm 0.01$ (mean and standard deviation over the 10 data sets)
which is relatively small compared to other networks. 
In the context of the interbank market 
a small $C$ is a reasonable result. While banks might be
interested in some diversification of interbank links keeping a link is
also costly. So if for instance two small banks have a link
with their head institution there is no reason for them to 
additionally open a link among themselves. 

\noindent
{\bf Average Shortest Path Length:}
We calculate the average path length for 
the three networks $A^l,A^a,A$  
with the Dijkstra  algorithm \cite{gib85} and find  
an average path length of 
$\bar{\ell}(A^l)=\bar{\ell}(A^a)=2.59 \pm 0.02$. 
Note the possibility that in a directed graph 
not all nodes can be reached and we restrict our statistics to 
the giant components of the directed graphs.  
The average path length 
in the (undirected) interbank connection network $A$ is 
$\bar{\ell}(A)= 2.26\pm0.03$. 
From these results the Austrian interbank 
network looks like a very small world with about three degrees of separation.
This result looks natural in the light of the community structure
described earlier. The two and three tier organization with 
head institutions and sub-institutions apparently leads 
to short interbank distances via the upper tier
of the banking system and thus to a low degree of separation.


Our analysis provides a first picture of the interbank network topology 
by studying a unique dataset for the Austrian interbank market. Even though 
it is small the Austrian interbank market is structurally
very similar to the interbank system in many European countries including
the large economies of Germany, France, and Italy.
We show that the liability (contract) size distribution follows 
a power law, which can be understood as being driven by underlying 
size and wealth distributions of the banks, which show similar 
power exponents.  
We find that the interbank network shows -- like many other realistic   
networks  -- power law dependencies in the
degree distributions. 
We could show that different scaling exponents relate to different 
network structure in different banking sectors within the total 
network. The scaling exponents by the agricultural banks (R) are very low, 
due to the hierarchical structure of this sector, while the other banks 
lead to scaling exponents also found in other complex real world networks. 
Regardless of the size of the scaling exponent, 
the existence of a power law is a strong indication 
for a stable network with respect to random bank 
defaults or even intentional attack \cite{barabasi00}. 
The interbank network shows a low
clustering coefficient, a result that mirrors the 
analysis of community structure which shows
a clear network pattern, where banks 
would first have links with their 
head institution, whereas these few head institutions 
hold links among each other. A consequence 
of this structure is that the interbank network is a 
small world with a very low ''degree of separation'' 
between any two nodes in the system. 
A further important message of this work 
is that it allows to exclude large classes of unrealistic types 
of networks for future modeling of interbank relations, 
which have so far been used in the literature. 

We thank J.D. Farmer for valuable comments to improve the paper and  
Haijun Zhou for making his dissimilarity index algorithm 
available to us. S.T. would like to thank the SFI and in 
particular J.D. Farmer for their great hospitality in the summer 
of 2003.


\begin{thebibliography}{99}
\bibitem{dormen03}
S.N. Dorogovtsev and J.F.F. Mendes, 
{\it Evolution of Networks: From Biological Nets to the Internet and WWW}, 
(Oxford University Press, 2003).

\bibitem{watts}
D.J. Watts, 
{\it Small Worlds:
The Dynamics of Networks between Order and Randomness}, 
(Princeton University Press, 1999).  
\bibitem{barabasi00}
R. Albert, H. Jeong, and A.-L. Barabasi, 
Nature  {\bf 406},  378-382 (2000).

\bibitem{hardeb00}
P. Hartmann and  O. DeBandt, 
European Central Bank Working Paper Nr. 35 (2000).

\bibitem{sum03} 
M. Summer,
Open Economies Review {\bf 1}, 43 (2003). 

\bibitem{algal00}
F. Allen and D. Gale,  
Journal of Political Economy {\bf 108}, 1 (2000).

\bibitem{freparroc00}
X. Freixas, L. Parigi, and J.C.  Rochet,  
Journal of Money Credit and Banking {\bf 32} (2000).

\bibitem{thurner03}
S. Thurner, R. Hanel, S. Pichler, 
Quantitative Finance {\bf 3}, 306-319 (2003). 

\bibitem{shemau98}
G. Sheldon and M. Maurer,  
Swiss Journal of Economics and Statistics {\bf 134}, 685 (1998).

\bibitem{uppwor02}
C. Upper and A. Worms, 
Deutsche Bundesbank, Dicussion paper 09 (2002). 

\bibitem{bligra97}
U. Blien and F. Graef, 
in I. Balderjahn, R. Mathar,  and  M. Schader, 
{\it Classification, Data Analysis, and Data Highways}
(Springer Verlag, 1997).

\bibitem{fanrajtsa97} 
S.C. Fang, J.R. Rajasekra and J. Tsao, 
{\it Entropy Optimization and Mathematical Programming} 
(Kluwer Academic Publishers, 1997). 

\bibitem{solomon}
S. Solomon, M. Levy, 
e-print: arXiv: cond-mat/0005416 (2000).

\bibitem{axtell} 
R.L. Axtell, 
Science {/bf 293} 1818 (2001).

\bibitem{wasserman94}
S. Wasserman and K. Faust, 
{\it Social Network Analysis: Methods and Applications}
(Cambridge University Press, 1994). 
 
\bibitem{ravasz02} 
E. Ravasz, A.L. Somera, D.A. Mongru, Z.N. Oltvai, and 
A.-L. Barabasi, 
Science {\bf 297}, 1551 (2001). 

\bibitem{freeman77} 
L.C. Freeman,
Sociometry {\bf 40}, 35 (1977). 

\bibitem{girvan02} 
M. Girvan and M.E.J. Newman, 
Proc. Natl. Acad. Sci. {\bf 99}, 7831 (2002). 

\bibitem{zhou03_b} 
H. Zhou, 
Phys. Rev. E (in print) 
e-print: arXiv:cond-mat/0302030 (2003). 

\bibitem{zhou03_a} 
H. Zhou, 
e-print: arXiv:physics/0302032 (2003). 

\bibitem{albert2000} 
R. Albert, A.-L. Barabasi,  
Phys. Rev. Lett.  {\bf 85}, 5234-5237 (2000).

\bibitem{liljeros2001} 
F. Liljeros, C.F. Edling, L.A.N. Amaral, H.E. Stanley, Y., Aberg,  
Nature {\bf 411}, 907 (2001).

\bibitem{albert99}
R. Albert, H. Jeong, A.-L. Barabasi, 
Nature {\bf 401}, 130 (1999).

\bibitem{barabasi2002}
A.-L. Barabasi, H. Jeong, Z. Neda, E. Ravasz, A. Schubert, T. Vicsek, 
Physica A {\bf 311}, 590 (2002).

\bibitem{ebel2002}
H. Ebel, L.I. Mielsch, S. Bornholdt, 
Phys. Rev. E {\bf 66}, 036103 (2002).

\bibitem{newman01}
M.E.J. Newman, 
Phys. Rev. E {\bf 64}, 016131 and 016132 (2001).

\bibitem{montoya2001}
J.M. Montoya, R.V. Sole, 
Santa Fe Institute working papers, 00-10-059 (2000). 

\bibitem{trusina2003}
A. Trusina, S. Maslov, P. Minnhagen, K. Sneppen, 
e-print: arXiv:cond-mat/0308339 (2003). 

\bibitem{gib85}
A. Gibbons, 
{\it Algorithmic Graph Theory} 
(Cambridge University Press, 1985).


\end{thebibliography}
\end{document}